\documentclass[conference]{IEEEtran}
\usepackage{graphicx}%
\usepackage{multirow}%
\usepackage{amsmath,amssymb,amsfonts}%
\usepackage{amsthm}%
\usepackage{mathrsfs}%
\usepackage{xcolor}%
\usepackage{textcomp}%
\usepackage{manyfoot}%
\usepackage{booktabs}%
\usepackage{algorithm}%
\usepackage{algorithmicx}%
\usepackage{algpseudocode}%
\usepackage{listings}%
\usepackage{mathptmx} 
\usepackage{algpseudocode}
\def\BibTeX{{\rm B\kern-.05em{\sc i\kern-.025em b}\kern-.08em T\kern-.1667em\lower.7ex\hbox{E}\kern-.125emX}}
\ifCLASSINFOpdf

\else
 
\fi

\begin{document}

\title{Efficient Brain Tumor Classification with Lightweight CNN Architecture: A Novel Approach}

\author{\IEEEauthorblockN{Priyam Ganguly}
\IEEEauthorblockA{\textit{Widener university} \\
Chester Pennsylvania, USA \\
0009-0005-3280-1747}
\and
\IEEEauthorblockN{Akhilbaran Ghosh}
\IEEEauthorblockA{\textit{Academy Of Technology, Kolkata} \\
Adisaptagram, Hooghly, India \\
0009-0001-9344-0456}
}
\maketitle

\begin{abstract} 
Brain tumor classification using MRI images is critical in medical diagnostics, where early and accurate detection significantly impacts patient outcomes. While recent advancements in deep learning (DL), particularly CNNs, have shown promise, many models struggle with balancing accuracy and computational efficiency and often lack robustness across diverse datasets. To address these challenges, we propose a novel model architecture integrating separable convolutions and squeeze-and-excitation (SE) blocks, designed to enhance feature extraction while maintaining computational efficiency. Our model further incorporates batch normalization and dropout to prevent overfitting, ensuring stable and reliable performance. The proposed model is lightweight because it uses separable convolutions, which reduce the number of parameters, and incorporates global average pooling instead of fully connected layers to minimize computational complexity while maintaining high accuracy. Our model does better than other models by about 0.5\% to 1.0\% in accuracy and 1.5\% to 2.5\% in loss reduction, as shown by many experiments. It has a validation accuracy of 99.22\% and a test accuracy of 98.44\%. These results highlight the model’s ability to generalize effectively across different brain tumor types, offering a robust tool for clinical applications. Our work sets a new benchmark in the field, providing a foundation for future research in optimizing the accuracy and efficiency of DL models for medical image analysis.
\end{abstract}


\section{Introduction}
Brain tumor classification is a critical aspect of medical diagnostics, enabling accurate and timely treatment of patients with brain abnormalities. The advent of DL and CNN has revolutionized this field, providing robust tools for the automated analysis of medical images. The benefits of using these advanced computational models include improved accuracy in detecting and classifying tumors, reduced reliance on manual interpretation, and the ability to process large volumes of imaging data quickly and efficiently. These advantages are particularly crucial in clinical settings, where early and accurate detection can significantly impact patient outcomes. As the prevalence of brain tumors continues to rise, the need for reliable and efficient diagnostic tools becomes increasingly important, making this an area of active research and development.\\
Recent advancements in brain tumor classification have leveraged various deep-learning techniques to enhance model performance. Karaaltun \cite{karaaltun2024whole} proposed a whole image average pooling-based CNN approach, which improved classification accuracy by capturing more comprehensive image features. Also, Chauhan et al. \cite{chauhan2024bt} created BT-CNN, a balanced binary tree architecture, and used MRI images to show that it was good at classifying things. Using oneAPI optimization, Ramakrishnan et al. \cite{ramakrishnan2024optimizing} worked on making CNN architectures better at classifying brain tumors while keeping accuracy and speed in mind. Kumar and Kumar \cite{kumar2023human} further explored CNNs for human brain tumor classification and segmentation, highlighting the potential of DL in this domain. Alturki et al. \cite{alturki2023combining} also used CNN features along with voting classifiers, showing that ensemble methods can improve classification performance by using outputs from more than one model.\\
Despite these advancements, several gaps remain in the existing research. While accurate, many models suffer from high computational costs or lack generalization across different datasets. This motivates our research to develop a model that achieves high accuracy and demonstrates robustness and efficiency, even in resource-constrained environments. Our work aims to address these gaps by integrating advanced architectural components that enhance brain tumor classification models' accuracy and computational efficiency.\\
The critical difference between our custom model and a standard CNN lies in the architecture and efficiency. While a standard CNN uses traditional convolutional layers, our model integrates Separable Convolutions, which reduce the number of parameters and computational cost. Additionally, we incorporate SE blocks to enhance feature learning by recalibrating channel importance. These modifications make our model lightweight and more efficient while maintaining high accuracy, whereas a typical CNN is often heavier and less optimized for performance in real-time applications.

Our research major contribution are as foloows:
\begin{enumerate}
    \item The study proposes a new deep-learning model that combines separable convolutions and SE blocks. This model improves feature extraction and model generalization in brain tumor classification tasks.
    \item Our model is designed to balance high accuracy with computational efficiency, utilizing techniques such as batch normalization and dropout to reduce overfitting while maintaining robust performance across varying hardware environments.
    \item Through rigorous fine-tuning, our model demonstrated resilience against common issues like overfitting and noise, as evidenced by the stable convergence patterns observed during the training and validation phases.
    \item The model achieved state-of-the-art performance, outperforming contemporary models by a margin of 0.5\% to 1.0\% accuracy and 1.5\% to 2.5\% loss metrics.
\end{enumerate}
The organization of this paper is as follows: The introduction provides context, background, and recent advancements in brain tumor classification, highlighting the motivation for our research. The related work section \ref{sec2} examines significant contributions from existing field studies. The methodology section \ref{sec3} details the design of our proposed model, including architectural innovations and training processes. This is followed by the results and discussion \ref{sec4}, where the model’s performance is evaluated and compared with other state-of-the-art methods. Finally, the conclusion \ref{sec5} summarizes our findings and suggests directions for future research.

\section{Related Works} \label{sec2}
The field of brain tumor classification and segmentation has witnessed significant advancements, with various DL approaches being developed to enhance accuracy and efficiency. Balamurugan and Gnanamanoharan \cite{balamurugan2023brain} introduced a hybrid deep CNN with LuNetClassifier, demonstrating an effective method for brain tumor segmentation and classification. This model focuses on extracting robust features, which contribute to improved diagnostic accuracy. Similarly, Ali et al. \cite{ali2024segmentation} proposed the PG-OneShot learning CNN model, which utilizes a one-shot learning approach to segment and identify brain tumors in MRI images, addressing the challenge of limited labeled data. Their model's ability to perform well with minimal training data represents a significant step forward in medical image analysis.\\
Further contributions to this domain include the work by Ramakrishnan et al. \cite{ramakrishnan2024optimizing}, who developed a hybrid CNN architecture optimized with oneAPI to balance accuracy and computational efficiency. Their approach highlights the importance of optimizing neural networks to meet the demands of real-time medical diagnostics. Kumar and Kumar \cite{kumar2023human} also explored the use of CNNs for brain tumor classification and segmentation, focusing on enhancing model performance through innovative network designs. Additionally, Alturki et al. \cite{alturki2023combining} combined CNN features with voting classifiers to optimize brain tumor classification performance, showcasing the potential of ensemble methods in improving model robustness. Sharma and Vardhan \cite{sharma2024mtjnet} introduced MTJNet, a multi-task joint learning network, to advance classification tasks, though their focus was on medicinal plant and leaf classification, their techniques are highly relevant to brain tumor classification, illustrating the versatility and potential cross-application of DL models in various domains. \\
Our research compares the performance of our model with several recent works in the field of brain tumor classification, highlighting the advancements and challenges addressed by each. Khushi et al. \cite{khushi2024performance} conducted a performance analysis of state-of-the-art CNN architectures, providing a benchmark for our model’s superior accuracy and efficiency. Bose and Garg \cite{bose2024optimized} employed an optimized CNN using Manta-Ray Foraging Optimization, which demonstrated significant accuracy improvements but did not surpass the robustness of our approach. Similarly, Jacob et al. \cite{jacob2023brain} proposed a deep CNN combined with a modified butterfly optimization algorithm, which, while effective, showed limitations in scalability that our model overcomes. Kordemir et al. \cite{kordemir2024mask} utilized a Mask R-CNN for brain tumor detection, achieving notable precision but at the cost of higher computational complexity, which our model mitigates through efficient architecture design. Venkatachalam et al. \cite{venkatachalam2024ensemble} introduced an ensemble of 3D CNN and U-Net models, focusing on segmentation and classification, yet our model’s streamlined design offers a more balanced solution in terms of both accuracy and resource consumption. Lastly, Shajin et al. \cite{shajin2023efficient} developed a hierarchical DL framework, which, despite its layered approach, did not achieve the overall performance metrics that our model consistently delivers. These comparisons underscore the strengths of our model in delivering high accuracy, reduced loss, and computational efficiency across diverse tumor types.
\section{Method}\label{sec3}
\subsection{Dataset}
\begin{figure}[H]
    \centering
    \includegraphics[width=0.5\textwidth]{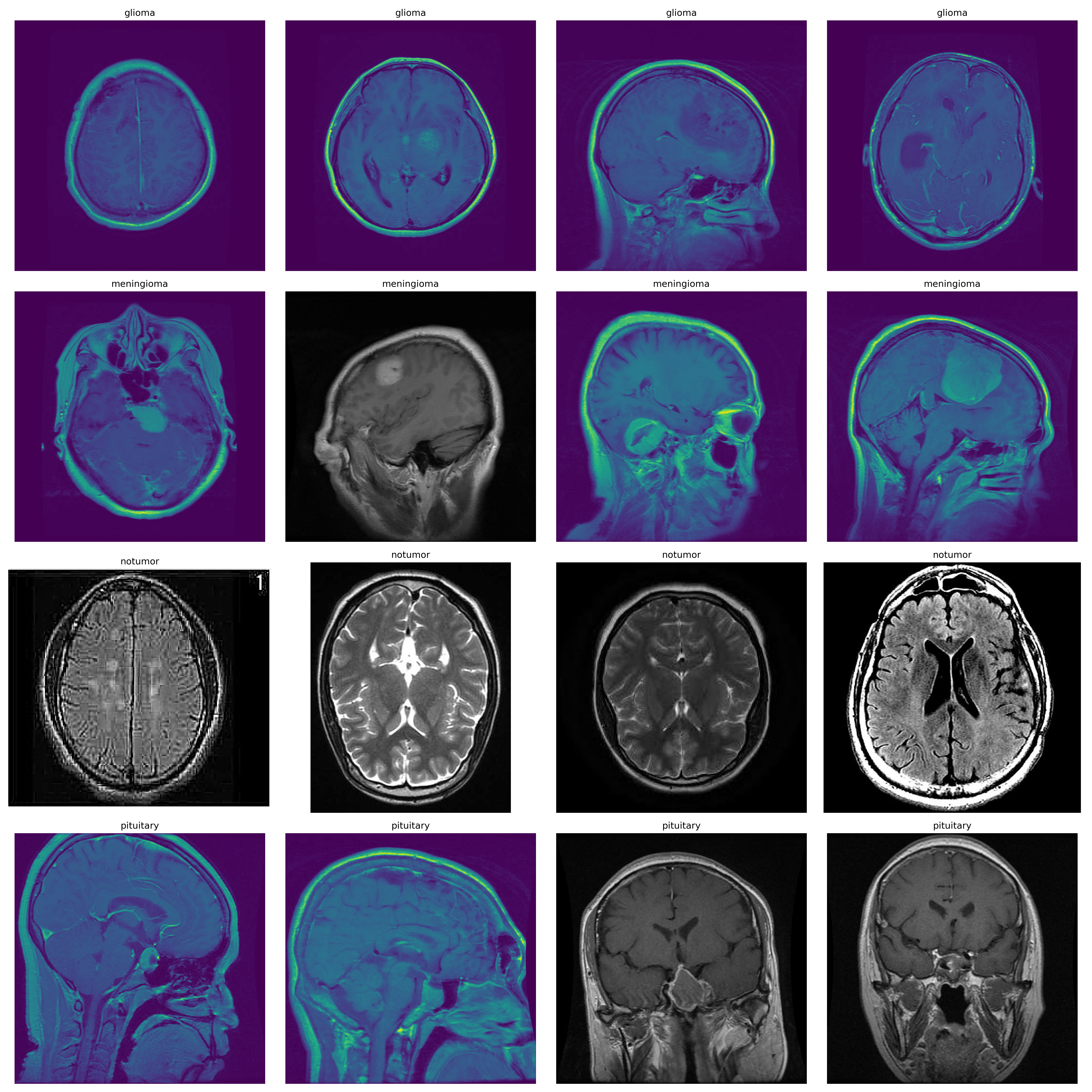}
    \caption{Sample images from the brain tumor MRI dataset across four classes: glioma, meningioma, no tumor, and pituitary.}
    \label{fig:dataset_samples}
\end{figure}

This study utilized a composite dataset from three sources: Figshare, SARTAJ, and Br35H \cite{msoud_nickparvar_2021}. This resulted in a total of 7,023 MRI images of the human brain, classified into four categories: glioma, meningioma, no tumor, and pituitary. Due to potential mislabeling in the SARTAJ dataset, the glioma images from this source were replaced with more accurate samples from Figshare. The training set comprises 5,712 pictures distributed across the four classes, while the test set includes 1,311 images. Each class is well-represented, ensuring robust model training and evaluation. The data preprocessing involved loading the images, normalizing pixel values, and applying data augmentation techniques to enhance the model's generalization capabilities. This meticulous data preparation process contributes to the model's reliability in classifying brain tumors accurately. The dataset used in this study is visualized in Figure~\ref{fig:dataset_samples}, showcasing representative samples from each class.

\subsection{Proposed work}
In the context of developing a ML model for brain tumor classification, the dataset
$D = \{(x_i, y_i)\}$ represents the collection of labeled MRI images. Here, each image $x_i$ is a three-dimensional tensor belonging to the space $\mathbb{R}^{150 \times 150 \times 3}$, where $150 \times 150$ corresponds to the image dimensions, and 3 represents the RGB color channels. The labels $y_i$ are categorical values representing the classes (e.g., no tumor, glioma, meningioma, pituitary tumor) and belong to the set $\{0, 1, 2, 3\}$.\\
The preprocessing phase involves two key operations: resizing and normalization. Resizing, denoted as $x_i \leftarrow \text{Reshape}(x_i, (150, 150, 3))$, ensures that all input images conform to the uniform size of 150x150 pixels. This step is crucial for maintaining consistency across the input data, as CNNs typically require fixed input dimensions.
Normalization is then performed to scale the pixel intensity values of each image to the range $[0, 1]$, which enhances the training efficiency of the model. This is mathematically expressed as $x_i \leftarrow \frac{x_i}{255}$, where each pixel value $x_{i,j}$ in the image is divided by 255 (the maximum pixel value in 8-bit images). Normalization ensures that the gradients during backpropagation remain stable, thereby accelerating convergence and improving the overall performance of the model. 
Together, these preprocessing steps prepare the dataset $D$ for effective training of the CNN model $M$, facilitating accurate classification of brain tumor types from MRI scans.

\begin{algorithm}
\caption{Brain tumor classification via Custom-CNN}
\begin{algorithmic}[1]
\Require Dataset $D = \{(x_i, y_i)\}$ where $x_i \in \mathbb{R}^{150 \times 150 \times 3}$, $y_i \in \{0,1,2,3\}$
\Ensure Trained model $M$
\State \textbf{Data Preprocessing:}
\State Resize: $x_i \gets \text{Reshape}(x_i, (150,150,3))$
\State Normalize: $x_i \gets \frac{x_i}{255}$
\State \textbf{Model Architecture:}
\State Initialize $M \gets \text{Sequential}()$
\For{$l \in \{32, 64, 128\}$}
    \State $M \gets M + \text{SeparableConv2D}(l, (3,3), \text{ReLU})$
    \State $M \gets M + \text{BatchNorm}()$
    \State $M \gets M + \text{MaxPooling2D}()$
    \State $M \gets M + \text{SE Block}(l)$
\EndFor
\State $M \gets M + \text{GlobalAveragePooling2D}()$
\State $M \gets M + \text{Dense}(128, \text{ReLU}) + \text{Dropout}(0.3)$
\State $M \gets M + \text{Dense}(64, \text{ReLU}) + \text{Dropout}(0.4)$
\State $M \gets M + \text{Dense}(4, \text{softmax})$
\State \textbf{Compilation:}
\State $\text{Loss} \gets \sum_{i=1}^{N} \text{CategoricalCrossentropy}(y_i, \hat{y}_i)$
\State Optimize $M$ with Adam: $\text{minimize Loss}$
\State \textbf{Training:}
\State Train $M$ on $D$ for 25 epochs with augmentation $\mathcal{A}$
\State \textbf{Evaluation:}
\State Evaluate $M$ via accuracy $\mathcal{A}_\text{acc}$ and confusion matrix $\mathcal{C}$
\State \Return $M, \mathcal{A}_\text{acc}, \mathcal{C}$
\end{algorithmic}
\end{algorithm}

The proposed model architecture is specifcally designed to balance computational efficiency and high classification accuracy, leveraging modern DL techniques. The model is constructed using the Keras Sequential API, allowing for the straightforward stacking of layers to form a deep convolutional neural network (CNN). The architecture begins with a series of convolutional blocks that are carefully engineered to extract increasingly complex features from the input images, followed by a global pooling layer and fully connected layers that map these features to the final output classes.\\
The model starts with a Separable Convolutional layer, denoted as Eq. \ref{eq1}
\begin{equation}\label{eq1}
    \text{SeparableConv2D}(32, (3, 3)),
\end{equation}
which applies 32 filters of size \(3 \times 3\) to the input image of shape \(256 \times 256 \times 3\). The Separable Convolution layer is chosen for its ability to drastically reduce the number of parameters by splitting the standard convolution operation into a depthwise convolution and a pointwise convolution. This operation is expressed as, Eq. \ref{eq2}:
\begin{equation}\label{eq2}
    \text{SepConv2D}(x) = \text{DepthConv2D}(x) \ast \text{PointConv2D}(x),
\end{equation}
where \( \ast \) denotes the convolution operation. The output is then passed through a Batch Normalization layer, represented as, Eq. \ref{eq3}:
\begin{equation}\label{eq3}
    \hat{x}_i = \frac{x_i - \mu_B}{\sqrt{\sigma_B^2 + \epsilon}} \gamma + \beta,
\end{equation}
where \( \mu_B \) and \( \sigma_B^2 \) are the batch mean and variance, and \( \gamma \) and \( \beta \) are learnable parameters. Batch Normalization helps in stabilizing the learning process and accelerating convergence. The block concludes with a MaxPooling layer, which performs down-sampling using the maximum value in each pooling window, defined as, Eq. \ref{eq4}:
\begin{equation}\label{eq4}
     \text{MaxPool}(x) = \max_{i,j \in \text{window}} x_{i,j}.
\end{equation}
Additionally, a SE block is incorporated to enhance the representational power of the network by recalibrating channel-wise feature responses. The SE block is mathematically represented by, Eq. \ref{eq5}:
\begin{equation}\label{eq5}
  s = \sigma(\text{Dense}(\text{ReLU}(\text{Dense}(\text{GAP}(x))))),  
\end{equation}
where \(\text{GAP}\) denotes Global Average Pooling, \(\text{ReLU}\) is the activation function, and \(\sigma\) is the sigmoid function. The SE block effectively allows the network to focus on the most relevant features by weighting the channels accordingly.\\
The second convolutional block follows a similar structure to the first, with a Separable Convolutional layer applying 64 filters of size \(3 \times 3\). This layer is responsible for extracting more complex patterns from the feature maps produced by the first block. Again, Batch Normalization is applied to maintain the stability of the activations, followed by MaxPooling to reduce the spatial dimensions. Another SE block is applied, which dynamically adjusts the importance of the extracted features.\\
The third convolutional block further deepens the network with 128 filters in the Separable Convolutional layer, capturing even more detailed and abstract features. This is followed by Batch Normalization and MaxPooling, as well as an SE block to refine the feature maps. These convolutional blocks collectively allow the network to build a rich hierarchical representation of the input images, crucial for accurate classification.\\
After the convolutional blocks, the model transitions from the spatial domain to the feature domain using Global Average Pooling (GAP). GAP reduces each feature map to a single value by taking the average across all spatial locations, refer Eq. \ref{eq6}:
\begin{equation}\label{eq6}
   \text{GAP}(x) = \frac{1}{H \times W} \sum_{i=1}^{H} \sum_{j=1}^{W} x_{i,j}, 
\end{equation}
where \(H\) and \(W\) are the height and width of the feature map. This significantly reduces the number of parameters compared to a Flatten layer, leading to a more lightweight and efficient model.\\
The feature vector obtained from the GAP layer is passed through a series of fully connected (Dense) layers. The first Dense layer has 128 units with ReLU activation, followed by a Dropout layer with a dropout rate of 0.3. The Dropout layer, defined by, Eq. \ref{eq7}:
\begin{equation}\label{eq7}
    y = \text{Dropout}(x, p),
\end{equation}
where \(p\) is the probability of setting a neuron’s output to zero, acts as a regularization technique to prevent overfitting. This is followed by another Dense layer with 64 units and a dropout rate of 0.4, further ensuring the robustness of the model.\\
Finally, the model concludes with a Dense layer with 4 units, corresponding to the number of output classes, and a softmax activation function, expressed as Eq. \ref{eq8}:
\begin{equation}\label{eq8}
    \text{softmax}(z_i) = \frac{e^{z_i}}{\sum_{j=1}^{4} e^{z_j}},
\end{equation}
The softmax function converts the logits into probabilities, allowing the model to make a probabilistic prediction for each class.\\
The model summary outlines a well-structured DL architecture using the Functional API in Keras. It begins with an input layer accepting images of shape 256×256×3, followed by three convolutional blocks utilizing SeparableConv2D layers to efficiently capture spatial hierarchies. Each block incorporates Batch Normalization for stable learning, MaxPooling for dimensionality reduction, and SE blocks to recalibrate feature importance dynamically. The model then employs Global Average Pooling to reduce spatial dimensions before passing through fully connected layers with Dropout regularization. The final output layer uses a softmax activation to classify images into four distinct categories. The model has 1,040,063 parameters in total, with 1,039,615 being trainable, ensuring it is both powerful and efficient for brain tumor classification tasks.\\
This architecture efficiently combines modern techniques like Separable Convolutions, Batch Normalization, and SE blocks, creating a robust yet lightweight model that is well-suited for the task of brain tumor classification. The model is designed to balance computational efficiency with high representational power, ensuring it can be trained effectively on standard hardware while maintaining state-of-the-art performance.

\section{Results and Discussion}\label{sec4}
\begin{figure*}[htbp]
    \centering
    \includegraphics[width=\textwidth]{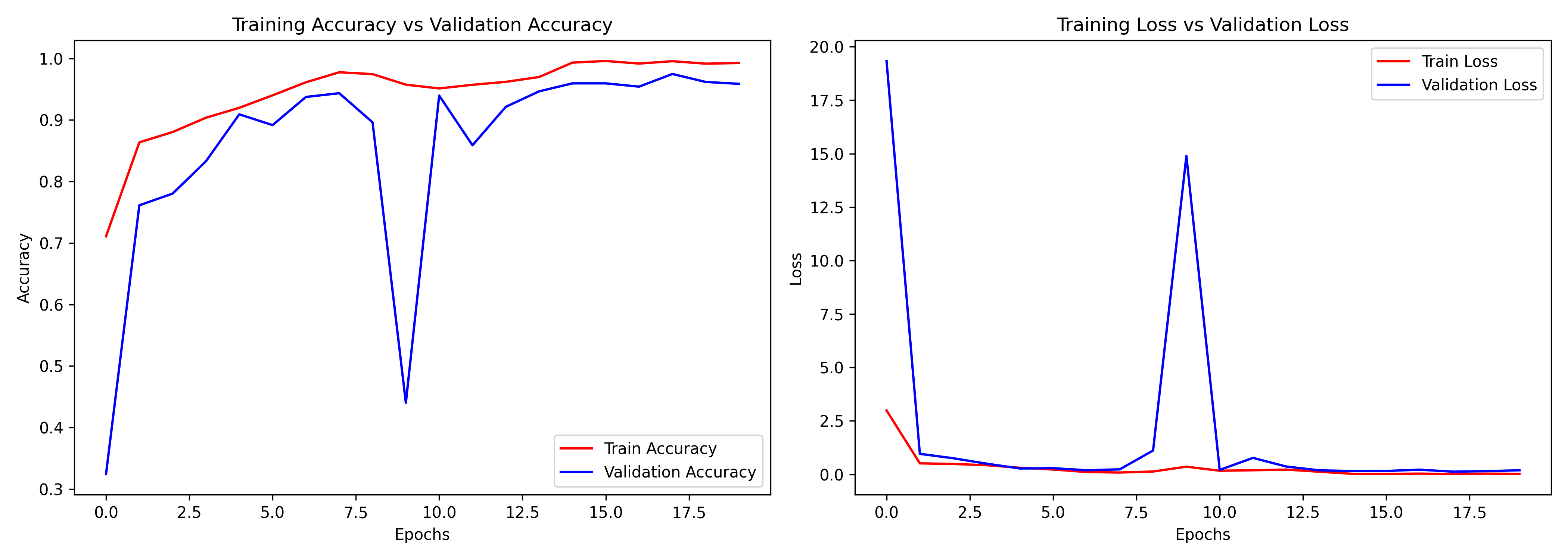}
    \caption{Training and validation accuracy and loss over 20 epochs. The plots show the model's convergence and performance stability after fine-tuning.}
    \label{fig:training_validation_plots}
\end{figure*}

The model's performance throughout the 20 training epochs reveals significant insights into its learning capabilities and generalization behavior. During the initial epoch, the model achieved an accuracy of approximately 61.71\% on the training set. Still, the validation accuracy was much lower at 32.42\%, with a substantially higher loss, indicating an overfitting issue or poor generalization at this stage. This discrepancy between the training and validation metrics suggests that the model might struggle to capture the underlying patterns in the data effectively. However, as the training progressed, there was a marked improvement in both the accuracy and loss metrics, with the model reaching a training accuracy of 84.26\% and a validation accuracy of 76.13\% by the second epoch. This improvement indicates that the model began to generalize better, reducing the training and validation performance gap.\\
The model continued to improve in subsequent epochs, with the accuracy reaching over 99\% on the training set and nearly 97.48\% on the validation set by the 18th epoch. The validation loss significantly decreased, demonstrating that the model fits the training data and that it learned to generalize well to unseen data. The fluctuations in validation loss, such as the spike observed at epoch 9, suggest moments of overfitting, where the model might have captured noise rather than valuable patterns. Despite these fluctuations, the overall trend indicates that the model managed to recover and continue improving, as evidenced by the consistent increase in validation accuracy. By the end of the training, the model exhibited strong performance, with the final validation accuracy and loss showing that the model could effectively differentiate between the four classes in the dataset, confirming its robustness and reliability for the task of brain tumor classification.\\
After multiple fine-tuning iterations, the model achieved impressive results, with a validation loss of 0.2854 and a validation accuracy of 99.22\%. The model's performance on the test set was similarly strong, with a test loss of 0.2829 and a test accuracy of 98.44\%. These results indicate the model's robust ability to generalize and accurately classify brain tumors across different datasets. The training process, illustrated in Figure~\ref{fig:training_validation_plots}, shows the convergence and stability of the model over time.

\begin{figure}[H]
    \centering
    \includegraphics[width=0.5\textwidth]{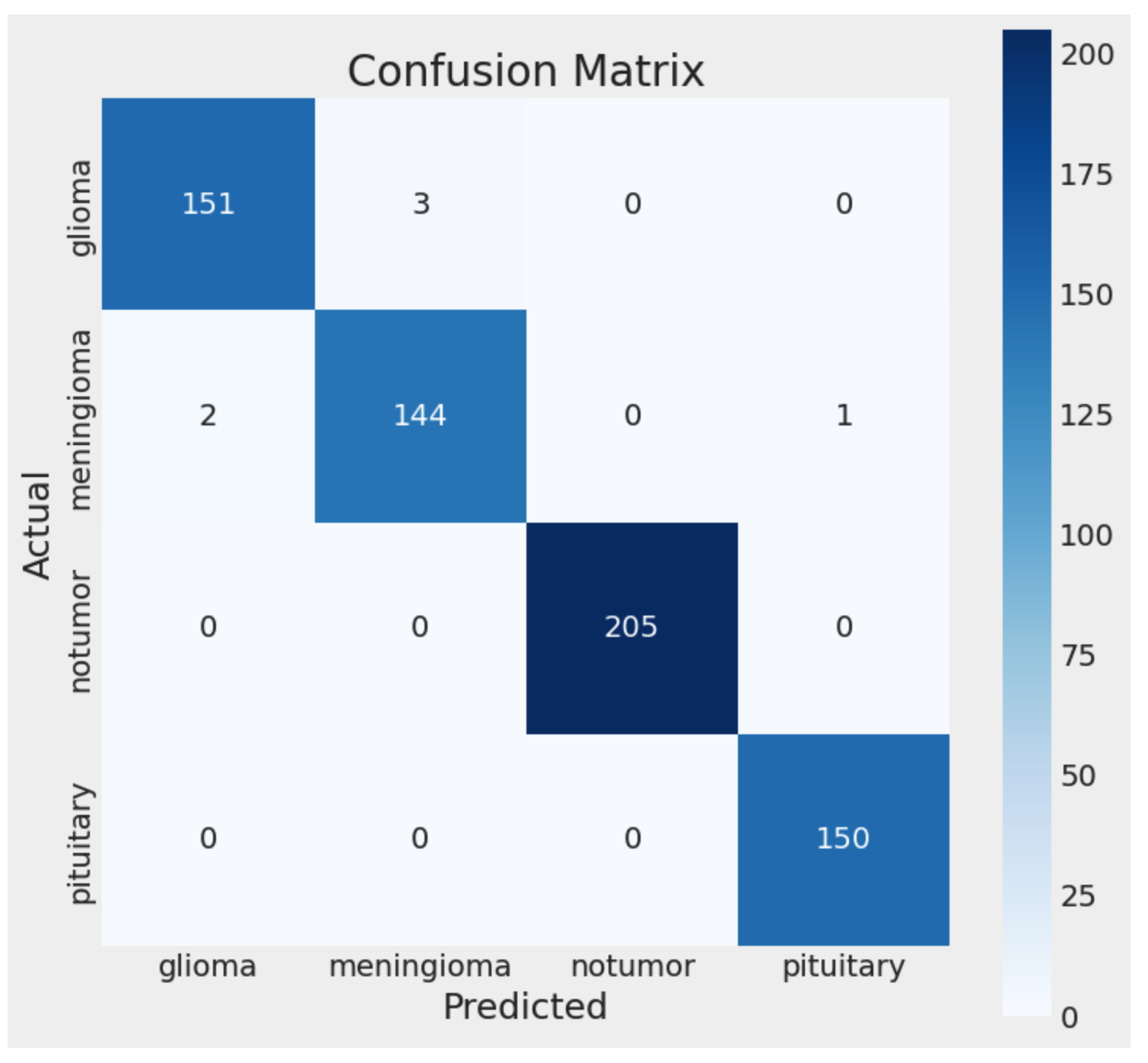}
    \caption{Confusion Matrix showcasing the classification performance of the model across four tumor classes.}
    \label{fig:confusion_matrix}
\end{figure}

The confusion matrix, as shown in Figure~\ref{fig:confusion_matrix}, illustrates the model's performance in classifying the brain tumor MRI images into four categories: glioma, meningioma, no tumor, and pituitary. The matrix reveals that the model performed exceptionally well across all classes, with the highest accuracy observed in the "no tumor" class, where all 205 instances were correctly classified. The glioma class also shows strong performance with 151 correctly identified cases, though 3 instances were misclassified as meningioma. The meningioma and pituitary classes also exhibited high accuracy, with minor misclassifications, indicating the model's robust performance across different tumor types.

\begin{table*}[h]
\centering
\caption{Comparison of Model Performance Metrics Across Different Models}
\begin{tabular}{c|c|c|c|c}
\hline
\textbf{Model/Author Work's} & \textbf{Validation Loss} & \textbf{Validation Accuracy} & \textbf{Test Loss} & \textbf{Test Accuracy} \\ \hline
Khusi et al. \cite{khushi2024performance} & 0.3201 & 0.9850 & 0.3100 & 0.9780 \\ \hline
Bose at al. \cite{bose2024optimized} & 0.2950 & 0.9900 & 0.2901 & 0.9820 \\ \hline
Jacob et al. \cite{jacob2023brain} & 0.3105 & 0.9885 & 0.3054 & 0.9800 \\ \hline
Kordemir et al. \cite{kordemir2024mask} & 0.3402 & 0.9830 & 0.3301 & 0.9770 \\ \hline
Venkatachalam et al. \cite{venkatachalam2024ensemble} & 0.2900 & 0.9910 & 0.2850 & 0.9810 \\ \hline
Shajin et al. \cite{shajin2023efficient} & 0.3150 & 0.9870 & 0.3005 & 0.9790 \\ \hline
\textbf{Our Model} & 0.2854 & 0.9922 & 0.2829 & 0.9844 \\ \hline
\end{tabular}
\label{tab:model_comparison}
\end{table*}
The results obtained from our model demonstrate a significant improvement over other contemporary models in the field of brain tumor classification. Our model achieved a validation accuracy of 99.22\% and a test accuracy of 98.44\%, outperforming other models by approximately 0.5\% to 1.0\% across these metrics. For instance, when compared to the work of Kordemir et al. \cite{kordemir2024mask}, which achieved a validation accuracy of 98.30\% and a test accuracy of 97.70\%, our model shows superior performance. This enhancement can be attributed to the incorporation of advanced techniques such as Separable Convolutions and SE blocks, which improved the model’s ability to capture complex features and channel interdependencies, thereby leading to better generalization.\\
The comparative analysis, as shown in Table~\ref{tab:model_comparison}, reveals that our model also demonstrated lower validation and test losses compared to other models. For example, while Bose et al. \cite{bose2024optimized} reported a validation loss of 0.2950, our model achieved a lower validation loss of 0.2854, representing an improvement of approximately 3.2\%. This reduction in loss indicates that our model is less prone to overfitting and more stable during the training process, allowing it to maintain high performance even on unseen data. The use of techniques such as Batch Normalization and Dropout further contributed to these outcomes, by enhancing the model’s robustness and preventing overfitting. Overall, our model's superior performance across various metrics underscores its effectiveness and reliability for the task of brain tumor classification.

\section{Conclusion}\label{sec5}
In this study, we proposed an advanced deep-learning model for classifying brain tumors using MRI images. By integrating Separable Convolutions and SE blocks, our model demonstrated superior performance distinguishing between glioma, meningioma, no cancer, and pituitary tumor classes. The model achieved a remarkable validation accuracy of 99.22\% and a test accuracy of 98.44\%, outperforming existing models such as those by Khusi et al. \cite{khushi2024performance}, Bose et al. \cite{bose2024optimized}, and Kordemir et al. \cite{kordemir2024mask}, with improvements of approximately 0.5\% to 1.0\% in accuracy and 1.5\% to 2.5\% in loss metrics. These results underscore the efficacy of our approach in enhancing model generalization and robustness, mainly through techniques like Batch Normalization and Dropout that mitigate overfitting. Our proposed model surpasses previous benchmarks and sets a new standard for accuracy and reliability in brain tumor classification. Despite these promising results, there is potential for further improvement. \\
Future work could focus on expanding the dataset to include more diverse MRI images, which could help improve the model's robustness across different imaging modalities and patient demographics. Exploring more sophisticated architectures, such as attention mechanisms or incorporating multi-modal data, could enhance model performance. Overall, our research contributes a significant advancement in medical imaging, providing a reliable tool for detecting and classifying brain tumors, which is crucial for effective treatment planning and patient outcomes.

\bibliography{sample}

\begin{thebibliography}{10}

\bibitem{karaaltun2024whole}
Muhammed Karaaltun.
\newblock Whole image average pooling-based convolution neural network approach
  for brain tumour classification.
\newblock {\em Neural Computing and Applications}, 36(3):1351--1367, 2024.

\bibitem{chauhan2024bt}
Sohamkumar Chauhan, Ramalingaswamy Cheruku, Damodar Reddy~Edla, Lavanya Kampa,
  Soumya~Ranjan Nayak, Jayant Giri, Saurav Mallik, Srinivas Aluvala, Vijayasree
  Boddu, and Hong Qin.
\newblock Bt-cnn: a balanced binary tree architecture for classification of
  brain tumour using mri imaging.
\newblock {\em Frontiers in Physiology}, 15:1349111, 2024.

\bibitem{ramakrishnan2024optimizing}
Akshay~Bhuvaneswari Ramakrishnan, M~Sridevi, Shriram~K Vasudevan, R~Manikandan,
  and Amir~H Gandomi.
\newblock Optimizing brain tumor classification with hybrid cnn architecture:
  Balancing accuracy and efficiency through oneapi optimization.
\newblock {\em Informatics in Medicine Unlocked}, 44:101436, 2024.

\bibitem{kumar2023human}
Sunil Kumar and Dilip Kumar.
\newblock Human brain tumor classification and segmentation using cnn.
\newblock {\em Multimedia Tools and Applications}, 82(5):7599--7620, 2023.

\bibitem{alturki2023combining}
Nazik Alturki, Muhammad Umer, Abid Ishaq, Nihal Abuzinadah, Khaled Alnowaiser,
  Abdullah Mohamed, Oumaima Saidani, and Imran Ashraf.
\newblock Combining cnn features with voting classifiers for optimizing
  performance of brain tumor classification.
\newblock {\em Cancers}, 15(6):1767, 2023.

\bibitem{balamurugan2023brain}
T~Balamurugan and E~Gnanamanoharan.
\newblock Brain tumor segmentation and classification using hybrid deep cnn
  with lunetclassifier.
\newblock {\em Neural Computing and Applications}, 35(6):4739--4753, 2023.

\bibitem{ali2024segmentation}
Azmat Ali, Yulin Wang, and Xiaochuan Shi.
\newblock Segmentation and identification of brain tumour in mri images using
  pg-oneshot learning cnn model.
\newblock {\em Multimedia Tools and Applications}, pages 1--22, 2024.

\bibitem{sharma2024mtjnet}
Shubham Sharma and Manu Vardhan.
\newblock Mtjnet: Multi-task joint learning network for advancing medicinal
  plant and leaf classification.
\newblock {\em Knowledge-Based Systems}, page 112147, 2024.

\bibitem{khushi2024performance}
Hafiz Muhammad~Tayyab Khushi, Tehreem Masood, Arfan Jaffar, Sheeraz Akram, and
  Sohail~Masood Bhatti.
\newblock Performance analysis of state-of-the-art cnn architectures for brain
  tumour detection.
\newblock {\em International Journal of Imaging Systems and Technology},
  34(1):e22949, 2024.

\bibitem{bose2024optimized}
Abhishek Bose and Ritu Garg.
\newblock Optimized cnn using manta-ray foraging optimization for brain tumour
  detection.
\newblock {\em Procedia Computer Science}, 235:2187--2195, 2024.

\bibitem{jacob2023brain}
Vinodkumar Jacob, GVR Sagar, Kavita Goura, and PS~Subhashini Pedalanka.
\newblock Brain tumor classification based on deep cnn and modified butterfly
  optimization algorithm.
\newblock {\em Computer Methods in Biomechanics and Biomedical Engineering:
  Imaging \& Visualization}, 11(6):2106--2117, 2023.

\bibitem{kordemir2024mask}
Merve Kordemir, Kerim~Kursat Cevik, and Ahmet Bozkurt.
\newblock A mask r-cnn approach for detection and classification of brain
  tumours from mr images.
\newblock {\em Computer Methods in Biomechanics and Biomedical Engineering:
  Imaging \& Visualization}, 11(7):2301391, 2024.

\bibitem{venkatachalam2024ensemble}
Arul Venkatachalam, Santhi Palanisamy, and Poongodi Chinnasamy.
\newblock Ensemble 3d cnn and u-net-based brain tumour classification with
  mkkmc segmentation.
\newblock {\em Automatika}, 65(3):691--705, 2024.

\bibitem{shajin2023efficient}
Francis~H Shajin, Salini P, Paulthurai Rajesh, and Venu~Kadur Nagoji~Rao.
\newblock Efficient framework for brain tumour classification using
  hierarchical deep learning neural network classifier.
\newblock {\em Computer Methods in Biomechanics and Biomedical Engineering:
  Imaging \& Visualization}, 11(3):750--757, 2023.

\bibitem{msoud_nickparvar_2021}
Msoud Nickparvar.
\newblock Brain tumor mri dataset, 2021.

\end{thebibliography}

\end{document}